\documentstyle[12pt,epsf]{article}
\def\J{$J/\psi$}

\def\P{$\psi'$}

\def\U{$\Upsilon$}

\def\q{q{\bar q}}

\def\be{\begin{equation}}
\def\ee{\end{equation}}

\def\lsim{\raise0.3ex\hbox{$<$\kern-0.75em\raise-1.1ex\hbox{$\sim$}}}
\def\gsim{\raise0.3ex\hbox{$>$\kern-0.75em\raise-1.1ex\hbox{$\sim$}}}

\def\PL{{ Phys.\ Lett.\ }}
\def\PR{{ Phys.\ Rev.\ }}

\def\PRL{{ Phys.\ Rev.\ Lett.\ }}

\def\ZP{{ Z.\ Phys.\ }}

\begin{document}

\noindent May 22, 2001~ \hfill BI-TP 2001/08

\vskip 1.5 cm

\centerline{\large{\bf String Breaking and
Quarkonium Dissociation}}

\medskip

\centerline{\large{\bf at Finite Temperatures}}

\vskip 1.0cm

\centerline{\bf S.\ Digal, P.\ Petreczky and H.\ Satz}

\bigskip

\centerline{Fakult\"at f\"ur Physik, Universit\"at Bielefeld}
\par
\centerline{D-33501 Bielefeld, Germany}

\vskip 1.0cm

\noindent

\centerline{\bf Abstract:}

\medskip

Recent lattice studies of string breaking in QCD with dynamical 
quarks determine the in-medium temperature dependence of the heavy 
quark potential. Comparing this to the binding energies of different 
quarkonium states, we check if these can decay into open charm/beauty
in a confined hadronic medium. Our studies indicate in particular that
the $\chi_c$ and the \P~dissociate into open charm below the 
deconfinement point.

\vskip1.5cm

The behavior of quarkonium states in a hot strongly interacting medium
was proposed as test for its confinement status \cite{Matsui}. 
In sufficiently hot deconfined matter, color screening will dissolve 
the binding of the heavy quark-antiquark pair; hence the
higher excited states with smaller binding energies and larger radii
will break up at lower temperatures. On a microscopic level, it was
argued that only a hot deconfined medium provides sufficiently hard
gluons to dissociate the tighly bound quarkonium states \cite{KS3};
this again leads to a dissociation hierarchy as function of the binding
energy. However, the arguments depend on a large binding energy, and it
is not clear to what extent they remain valid for higher excited states.
In particular, the \P~has a mass only 50 MeV below the open charm threshold,
to that it should be readily dissociated in any (confined or deconfined)
medium. It is thus of considerable interest to study possible 
collective mechanisms for quarkonium dissociation in a confined
medium.

\par

The study of string breaking at finite temperatures provides a new way 
to address this problem. String breaking will presumably occur when
the string potential reaches twice the mass of the lowest light-heavy 
meson \cite{Wittig}. Hence if a charmonium state, stable under strong 
interactions at $T=0$, is to be broken up in a confining environment with
$T>0$, the gap between the mass of the state and the open charm threshold 
has to be 
overcome, so that the decay into a $D^+D^-$ pair becomes energetically 
favorable. This can occur if the mass of the open charm meson decreases 
with temperature faster than the charmonium mass. Lattice studies of the 
heavy quark potential \cite{DeTar,Peikert} now determine the temperature 
dependence of the string breaking threshold. The investigation of the 
suppression of charmonium production below the deconfinement point thus 
becomes possible; the same arguments apply of course to bottomonium 
production as well. The aim of this paper is to study such hadronic 
in-medium decay 
dissociation of charmonium and bottomonium states.

\par

In the absence of any medium, charmonium and bottomonium spectra are quite
well described by non-relativistic potential theory
\cite{Cornell}-\cite{Eichten}. The basis for this is given by the
Schr\"odinger equation
\be
\left[ 2m_a + {1\over m_a}\nabla^2 + V_1(r) \right] \Phi_i^a = M_i^a
\Phi_i^a,
\label{1}
\ee
where $a=c,b$ specifies charm or bottom quarks, $i$ denotes the
quarkonium state in question, and the reduced radius $r$ is the
separation of the two heavy quarks.
The confining color singlet potential can be parametrized as
\be
V_1(r) = \sigma~r - {\alpha \over r}
\label{2}
\ee
in terms of the string tension $\sigma$ and a $1/r$ contribution
containing both Coulombic and transverse string effects \cite{Cornell};
for other forms, see \cite{Eichten}. It is evident that Eq.\ (\ref{2})
is a `quenched' idealization, since in full QCD the string connecting
the two heavy quarks will break when its energy surpasses that of two
dressed light quarks. Hence a more realistic form would be
\begin{eqnarray}
V_1(r) &=& \sigma~r - {\alpha \over r} ~~~~\forall ~~r~\lsim~r_h,
\nonumber\\
&=&E_s  ~~~~~~~~~~~\forall~~r~\gsim~r_h,
\label{3}
\end{eqnarray}
where $r_h \simeq E_s/\sigma$ indicates the typical hadronic
confinement scale and $E_s$ the energy needed to break the string
between the heavy quarks. Since this energy is related to the energy
required to bring a virtual light $\q$ pair on-shell as a hadron, we
expect it to be related to a chiral symmetry breaking constituent
quark mass, independent of the masses of the heavy quarks. From
Eq.\ (\ref{1}), the energy needed to break up quarkonium state $(a,i)$
is
\be
E_{a,i} = 2m_a + E_s^a - M_i^a .
\label{4}
\ee
To achieve a break-up, we have to overcome the energy difference between
the quarkonium state and two open charm or beauty mesons, so that
\be
E_{c,i} = 2M_D - M_i^c ~~~,~~~E_{b,i} = 2M_B - M_i^b,
\label{5}
\ee
with $M_D$ and $M_B$ denoting the lightest open charm and beauty mesons,
respectively. Using $m_c \simeq 1.3$ GeV and $m_b\simeq 4.7$ GeV as
indicative heavy quark mass values \cite{Jacobs,Bali}, we get from these two
equations
\be
E_s^c = 2(M_D-m_c) \simeq 1.1  ~ {\rm GeV}
\label{6}
\ee
and
\be
E_s^b = 2(M_B-m_b) \simeq 1.1 ~ {\rm GeV},
\label{7}
\ee
for $c \bar c$ and $b \bar b$ states, respectively. The fact that the 
two values agree is reassuring in two ways: first, it indicates that 
indeed the chiral properties of the QCD vacuum determine $E_s$, not 
the masses of the heavy quarks. Secondly, with $\sigma \simeq 0.19$ GeV$^2$
\cite{Jacobs,Bali}, the break-up occurs for $r_h \simeq 1/\sigma \simeq 1$
fm, in accord with the expected range of confining forces.

\par

In the presence of a medium, we expect the break-up to become easier
with increasing temperature, since presumably the effective constituent 
quark mass as well as the string tension will decrease. Previously
\cite{MTM,K-S}, the evolution of quarkonium binding with temperature
was studied on the basis of a screened form of the Cornell potential
(\ref{2}); the dissociation values of the screening mass, defined by
the divergence points of the corresponding bound state radii, were then
related to the temperature through quenched finite temperature lattice
results. Such a procedure can at best give some first idea of the
dissociation pattern; in particular, it can only give upper limits for
the existence of higher excited states. For a more realistic treatment, 
the heavy quark potential must be determined directly in lattice studies 
of full QCD. This will then not only allow us to study the effect of 
color screening in a deconfined medium, but also the evolution of the 
gap between the bound state mass and the open charm/beauty threshold. 
Such studies will indicate if and for what states the dissociation can 
occur already in a confined medium and at the same time provide the 
associated dissociation temperatures.

\par

In finite temperature lattice QCD, the temperature behavior of the
heavy quark potential $V(T,r)$ is obtained from Polyakov loop
correlations,
\be
- T~\ln <L(0)L^+(r)> = V(T,r) + C.
\label{8}
\ee
The constant $C$ includes both the cut-off dependent self-energy and 
the entropy contributions $-TS$ (for $T\not=0$, $-T \ln <L(0)L^+(r)>$ 
is actually the free energy of the $Q \bar Q$ pair and not the static 
energy, as in the zero temperature case \cite{Peikert}). The constant 
$C$ can be fixed by requiring that at very short
distances, $r << T^{-1}$, the potential has the zero temperature
form, since in the limit $r \to 0$, the effects of the medium
should become negligible. Present lattice calculations are perhaps 
not yet precise enough to reach sufficiently small $r$; nevertheless, 
they should provide some reasonable indication of the behavior of 
the potential.

\par

\begin{figure}
\epsfxsize=11cm
\epsfysize=9cm
\centerline{\epsffile{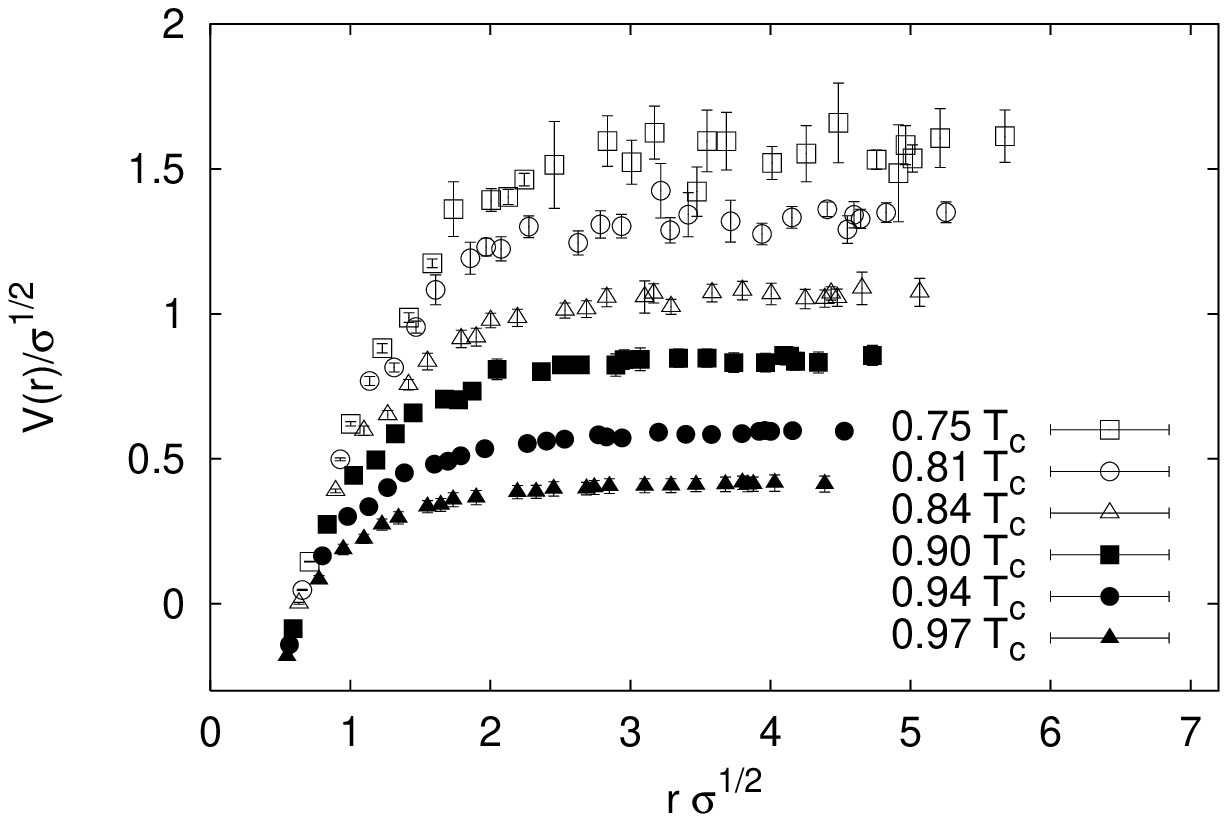}}
\caption{The heavy quark potential at different temperatures,
normalized to the Cornell potential at short distance.}
\label{F(r)}
\end{figure}

\begin{figure}
\epsfxsize=11cm
\epsfysize=9cm
\centerline{\epsffile{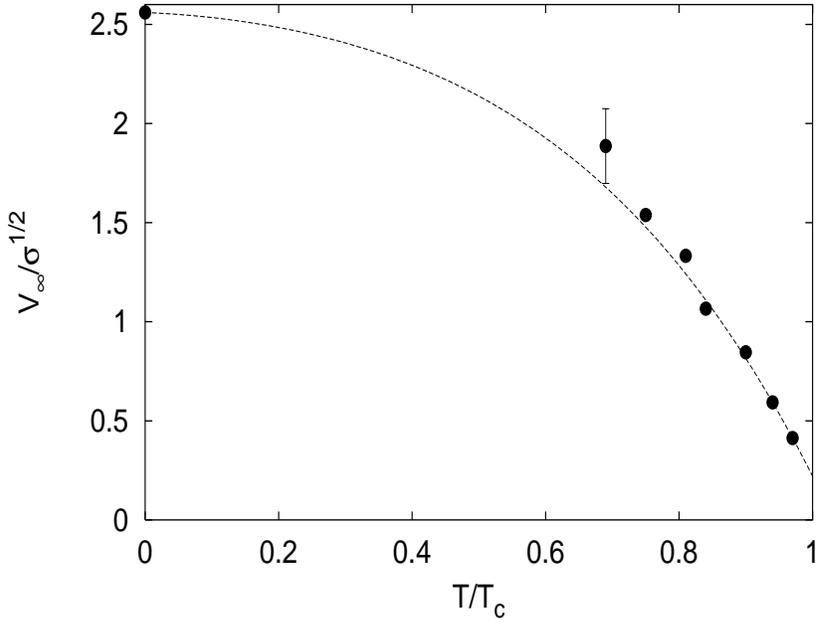}}
\caption{The temperature-dependence of the string-breaking plateau of the
heavy quark potential; the zero temperature point is obtained from Eqs.\ 
(\ref{6},\ref{7}).}
\label{F(T)}
\end{figure}

The free energy (\ref{8}) was recently studied in considerable detail
on $16^3 \times 4$ lattices, for three species of light quarks,
using an improved gauge and staggered fermion action \cite{Peikert}. Some
representative temperature curves of the resulting potential in
units of $\sqrt{\sigma}$
(\ref{8}) are shown in Fig.\ \ref{F(r)}. The normalization constant $C$ 
was fixed here by requiring the potentials to agree with the zero
temperature Cornell potential (\ref{2}) at $r=1/(4T)$. For this 
we have taken $\alpha=0.4$ in Eq.\ (\ref{2}), which is the value 
estimated in 3-flavor QCD \cite{Bali}; the value of $\sigma$ is irrelevant
if $V(r,T)$ and $r$ are measured in units of $\sqrt{\sigma}$. It is seen that
beyond a certain separation distance, the potential of the heavy quark
system reaches a constant value $V_{\infty}(T)$, corresponding to string 
breaking and hence the formation of an ``open charm"  
$Q{\bar q}-{\bar Q}q$ pair. The results shown in Fig.\ \ref{F(r)} were
obtained for quark mass $m_q=0.4~T$; however, a more detailed study of
the quark mass dependence shows that this value is already
very close to the chiral limit \cite{Peikert}. The lattice results 
thus provide a reasonable estimate of the temperature behavior of
$V_{\infty}(T)$. It is shown in Fig.\ \ref{F(T)}, including the value 
$V_{\infty}(0) \simeq E_s \simeq 1.1$ GeV from Eqs.\ (\ref{6},\ref{7}).
We note that in the vicinity of the deconfinement point $T=T_c$, the potential 
drops quite sharply, as expected.

\par

As mentioned, the potential form shown in Figs.\ \ref{F(r)} and
\ref{F(T)} was obtained by normalizing the lattice results to the 
zero temperature Cornell potential (\ref{2}) at the
reference point $r=1/(4T)$; the present lattice size does not
permit smaller $r$. For $T=T_c$, the normalization point corresponds to
$r \simeq 0.3$ fm; however, then deviations from the Cornell potential 
set in at essentially this distance, so that for
$T \to T_c$, the normalization becomes unreliable. In general, medium
effects, such as a temperature dependence of the string tension $\sigma$, 
may begin to affect the static potential at this distance, and as a 
consequence, also the masses of quarkonia states may become 
temperature-dependent \cite{Vogt}. To check the potential form used for 
normalization, we therefore want to consider a temperature-dependent
string potential instead of Eq.\ (\ref{2}). Such a potential
was found to describe quenched lattice results on the heavy-quark
potential for $T<T_c$ quite well, if it has the form \cite{Gao,Kacz}
\begin{eqnarray}
V_{string}(r,T)&=&(\alpha-{1\over 6} {\rm arctan}(2 r T) )~{1\over r}+
\nonumber\\
&&
(\sigma(T)-{\pi T^2\over 3}-{2 T^2\over 3} {\rm arctan}{1\over r T})~r+
\nonumber\\
&&
{1\over 2} \ln(1+4 r^2 T^2).
\label{vTdep}
\end{eqnarray}
\noindent
Normalizing the Polyakov loop correlator (\ref{8}) at $r=1/(4T)$ with 
Eq.\ (\ref{vTdep}), we thus obtain what might be a more reliable estimate 
of the plateau $V_{\infty}(T)$ than with the $T=0$ form (\ref{2}). 
In Eq.\ (\ref{vTdep}), we again set $\alpha=0.4$ \cite{Bali} and take 
$\sigma(T)$ as given in \cite{Kacz}. It turns out, however,
 that the two forms of
short distance behavior resulting from Eqs.\ (\ref{2}) and (\ref{vTdep})
are practically identical, so that the normalisation is in fact not affected
by the in-medium modifications at larger distances. 

\par

\begin{figure}
\epsfxsize=11cm
\epsfysize=9cm
\centerline{\epsffile{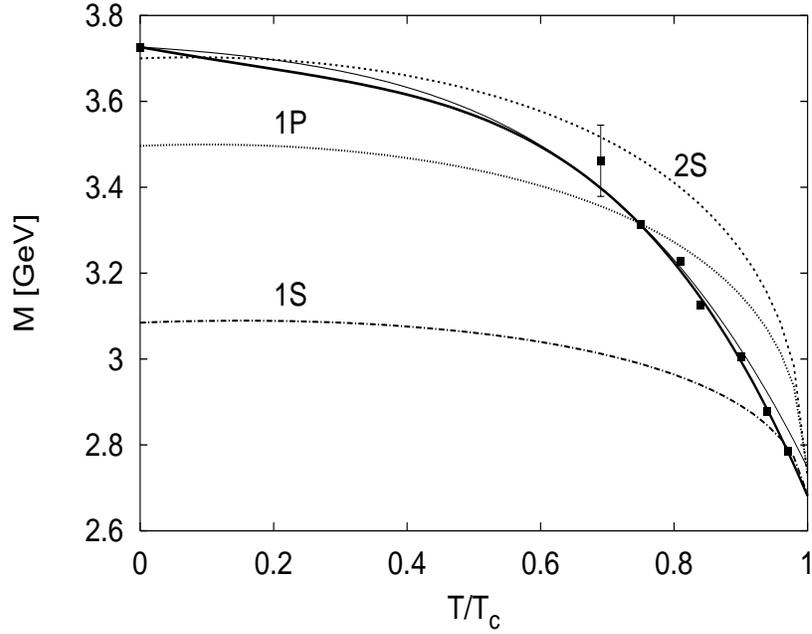}}
\caption{The open charm threshold and the masses of different charmonium 
states versus temperature. The thick solid line is the open charm threshold
obtained by normalizing at $rT=1/4$, the thin solid line at $rT=\sqrt{2}/4$.}
\label{psi}
\end{figure}

\begin{figure}
\epsfxsize=11cm
\epsfysize=9cm
\centerline{\epsffile{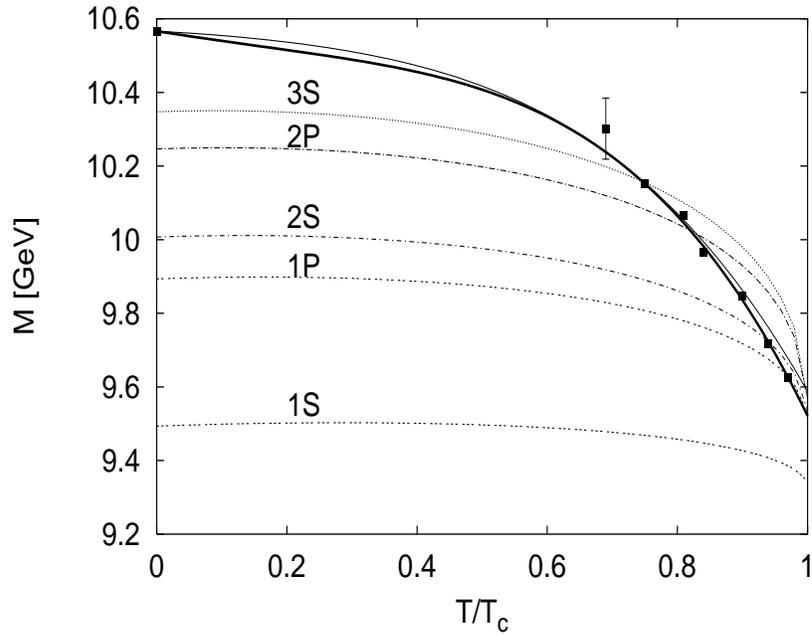}}
\caption{The open beauty threshold and the masses of
different bottomonium states versus the temperature.
The thick solid line is the open beauty threshold
obtained by normalizing
at $rT=1/4$, the thin solid line at $rT=\sqrt{2}/4$.
}
\label{upsi}
\end{figure}

To consider further possible uncertainties of the normalisation procedure,
we have also normalized the Polyakov loop correlator at the next smallest
distance $r=\sqrt{2}/(4T)$. The resulting two forms of $V_{\infty}(T)$ 
are shown in both Fig.\ \ref{psi} and Fig.\ \ref{upsi}. The difference 
between the two curves of $V_{\infty}(T)$ provides an estimate of 
the normalization error. Except for the region very near $T=T_c$, 
the uncertainty is seen to be quite small.

\par

Next, we then estimate the quarkonium masses based on (\ref{vTdep}). 
In the zero temperature limit, Eq.\ (\ref{vTdep}) reduces to Eq.\ (\ref{2}); 
this describes the quarkonium results quite well, if relativistic corrections 
to the potential are taken into account \cite{Bali}. The temperature 
dependence of the relativistic correction to the static potential is not 
known. For bottomonium, the relativistic corrections are small; for
charmonium, they may have to be included. On the other hand, we have 
found that a reasonably good description can be obtained with the 
c-quark mass $m_c=1.3$ GeV (in \cite{Bali}, the value $1.38$ was used). 
Thus zero temperature quarkonium spectroscopy can be reproduced quite well
with $\alpha=0.4$, $\sqrt{\sigma}=0.44$ GeV, $m_b=4.72$ and $m_c=1.3$ GeV.
These values were used to predict the finite temperature quarkonium masses
in our present study. In Table 1 we list all states, together with the 
corresponding zero temperature dissociation energies, and in Figs.\ \ref{psi} 
and \ref{upsi} we include the resulting temperature dependence of the
quarkonium masses, in order to determine their survival patterns. It is 
seen that \J~, 
\U~, and $\chi_b$ states are likely to survive up to temperatures 
$T \geq T_c$ and thus are out of the region we are able to consider here. 
The \U' state dissociates in a temperature region in which the value
of the open beauty threshold suffers from large uncertainties in the 
normalization of the Polyakov loop correlator. Thus for this state we can 
only give an lower limit for the dissociation.
On the other hand, the \P, $\chi_c$, $\chi_b'$ and \U'' masses intersect 
the string breaking curve noticably below $T_c$; the 
intersection values are listed in Table 1. 
Their dissociation occurs through hadronic decay, made possible because 
in-medium effects have effectively reduced the constituent quark 
mass enough to lower the open charm or beauty threshold below the 
mass values of the respective states. Hence they simply break up into open 
charm or beauty mesons; this is not a signal for deconfinement, but rather 
indicates the approach of chiral symmetry restoration and the resulting
decrease of the constituent quark mass. To probe the medium 
in a deconfined state, we have to consider the more tightly bound states \J,
$\chi_b$ and \U. 

\par

Evidently these conclusions differ from considerations of $\chi_c$
dissociation as possible signal for the onset of deconfinement
\cite{HS}. It should be noted, however, that previous screening studies
\cite{K-S} really could conclude only that \P~and $\chi_c$ cannot
exist for $T \geq T_c$; a prior in-medium dissociation of the type
obtained here was not excluded, but it could be addressed only once the
temperature dependence of the heavy quark potential and thus of string 
breaking became
accessible.

\begin{center}
\begin{tabular}{|c||c|c|c||c|c|c|c|c|}
\hline
&  &  & & & & & & \\
 state & \J & $\chi_c$ & \P & \U & $\chi_b$ & \U' & $\chi_b'$ & \U'' \\
&  &  & & & & & & \\
\hline
\hline
&  &  & & & & & & \\
$E_s^i$~[GeV] & 0.64 & 0.20 & 0.05 & 1.10 & 0.67 & 0.54 & 0.31 &
0.20 \\
&  &  & & & & & & \\
\hline
\hline
&  &  & & & & & & \\
$T_d/T_c$ & -  & 0.74 & 0.1 - 0.2 & - & - & $\gsim$~0.93  & 0.83 & 0.74 \\
&  &  & & & & & & \\
\hline
\end{tabular}\end{center}

\centerline{Table 1: Quarkonium Dissociation by String Breaking}

\bigskip

In summary, lattice studies of the heavy quark potential in QCD
with dynamical quarks indicate that there are two collective forms 
for quarkonium dissociation in hot matter. For $T < T_c$, hadronic
in-medium effects, related to the approach to chiral symmetry 
restoration and to the decrease of the string tension, lead to 
a reduction of the string-breaking threshold. This in turn results 
in the dissociation of the higher excited quarkonium
states (\P, $\chi_c$, $\chi'_b$, \U'') through decay into open
charm or beauty. The more tightly bound lower states (\J, \U, 
$\chi_b$ and perhaps also the \U') survive up to or beyond $T_c$. Their
dissociation will thus occur through color screening in a deconfined
medium; the analysis of this is carried out in \cite{D-P}, based on 
the corresponding QCD lattice studies for $T\geq T_c$ \cite{Peikert}.

\bigskip

\noindent
{\bf Acknowledgements}

\medskip

It is a pleasure to thank F. Karsch and E. Laermann for numerous 
helpful discussions. The financial support from DFG under grant
Ka 1198/4-1 and from BMFB under grant 06 BI 902 is gratefully 
acknowledged.

\newpage

\end{document}